\documentclass[12pt]{article}  

\usepackage{graphicx,amsmath}
\usepackage{units}

\newlength{\dinwidth}
\newlength{\dinmargin}
\def\lapproxeq{\lower .7ex\hbox{$\;\stackrel{\textstyle
<}{\sim}\;$}}
\def\gapproxeq{\lower .7ex\hbox{$\;\stackrel{\textstyle
>}{\sim}\;$}}
\def\gtrsim{\lower .7ex\hbox{$\;\stackrel{\textstyle
>}{\sim}\;$}}
\def\lesim{\lower .7ex\hbox{$\;\stackrel{\textstyle
<}{\sim}\;$}}

\def\be{\begin{equation}}
\def\ee{\end{equation}}
\def\bea{\begin{eqnarray}}
\def\eea{\end{eqnarray}}

\def\GeV{\rm GeV}

\begin{document}

\begin{flushright}                                                    
IPPP/10/84  \\
DCPT/10/168 \\                                                    
\today \\                                                    
\end{flushright} 

\vspace*{0.5cm}

\begin{center}
{\Large \bf An independent estimate of the triple-Pomeron coupling}

\vspace*{1cm}
                                                   
E.G. de Oliveira$^{a}$, A.D. Martin$^a$ and M.G. Ryskin$^{a,b}$ \\                                                    
                                                   
\vspace*{0.5cm}                                                    
$^a$ Institute for Particle Physics Phenomenology, University of Durham, Durham, DH1 3LE \\                                                   
$^b$ Petersburg Nuclear Physics Institute, Gatchina, St.~Petersburg, 188300, Russia

\vspace*{1cm}                                                    
                                                    
\begin{abstract}                                                    

The value of the important triple-Pomeron coupling is estimated by an unorthodox procedure using the known diffractive parton distribution functions. The result is $g_{3P}\simeq 0.2g_N$, where $g_N$ is the Pomeron-nucleon coupling. This is in excellent agreement with an independent determination, $g_{3P}\simeq 0.2g_N$, previously obtained by analysing the available data in the triple-Regge region with absorptive effects taken into account.

\end{abstract}                                                        
\vspace*{0.5cm}                                                    
                                                    
\end{center}

\vspace*{0.3cm}

`Soft' $pp$ interactions are known to have large absorptive corrections \cite{KMR}. Also, there is a significant contribution to the $pp$ cross section from high mass diffractive dissociation of the proton \cite{KMR}. Both of these effects are driven by the size of the triple-Pomeron coupling, $g_{3P}$. The  existing determination of this important parameter is obtained by analysing data in the triple-Regge region \cite{LKMR}. Here, to confirm the previous value, we use diffractive deep inelastic data to provide an alternative estimate of $g_{3P}$. 

\begin{figure} 
\begin{center}
\label{fig:A1}
\end{center}
\end{figure}

We begin by noting that, in Regge Theory, the interaction of two high-energy particles is described by Pomeron exchange. Here, we consider high-energy $pp$ (and $\bar{p}p$) collisions. Physically the Pomeron is represented by a ladder-type diagram.  Cutting the ladder diagram yields a `comb' of secondary particles, which correspond to inelastic interactions. The present Monte Carlo simulations are based on partons; the analogous chains of partons are generated as a semi-hard parton-parton collision supplemented by initial parton showers. These parton showers reproduce the DGLAP evolution leading to the final active partons, with $x_1$ and $x_2$ in Fig.~\ref{fig:A1}. Hence the inelastic cross section, which in terms of Regge theory corresponds to Pomeron exchange, is of the form
\be
\sigma^{\rm ND}_{pp}~=~\int dp_T^2dx_1dx_2~g(x_1,p_T^2)g(x_2,p_T^2)~\frac{d\hat{\sigma}}{dp_T^2}.
\label{eq:sig}
\ee
\begin{figure} 
\begin{center}
\includegraphics[height=6cm]{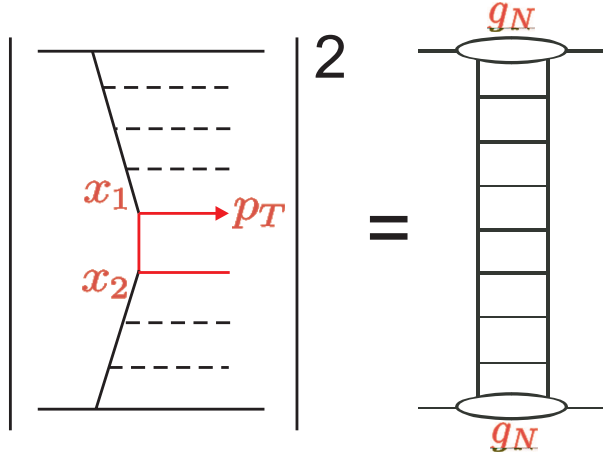}
\caption{\sf A diagram showing the inelastic $pp$ cross section driven by Pomeron exchange.}
\label{fig:A1}
\end{center}
\end{figure}
This expression is written simply in terms of gluons, but a summation over all parton pairs is implied. Formally, the integration over the transverse momenta, $p_T$, of the outgoing partons of the so-called `semi-hard' process, is infrared divergent as $p_T \to 0$. In the PYTHIA-8.1 Monte Carlo \cite{Pythia} this divergency is removed by replacing the $1/p_T^2$ propagator by $1/(p_T^2+p^2_{\rm min})$,
where
\be
p_{\rm min}~=~p_0\left(\frac{s}{s_0}\right)^{0.12},
\label{eq:para}
\ee
with the normalization $s_0=(1800~\GeV)^2$. The values $p_0=2.15 $ GeV and the exponent 0.12 were tuned\footnote{The tuning was done with CTEQ5L set of partons \cite{CTEQ5L}, which we therefore use throughout.} to describe the available {\it inclusive} data and correlations up to Tevatron energies.

\begin{figure} 
\begin{center}
\includegraphics[height=8cm]{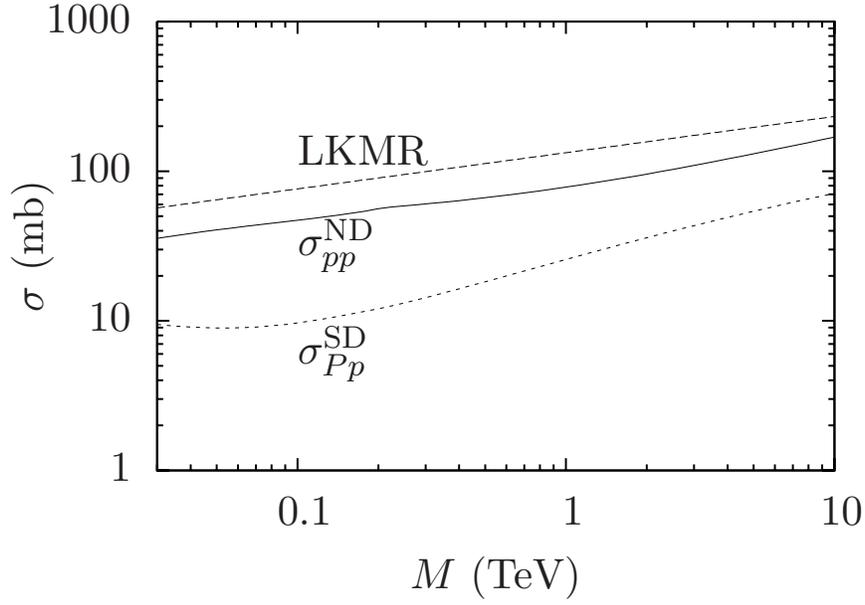}
\caption{\sf The results for the one-Pomeron exchange inelastic $pp$ and $Pp$ cross sections, which are respectively denoted $\sigma_{pp}^{\rm ND}$ (middle curve) and $\sigma_{Pp}^{\rm SD}$ (lower curve). The mass of the final hadronic system $M=\sqrt{s}$ for the $pp$ case and $M=\sqrt{x_P s}$ for the $Pp$ case. In this plot we take $x_P=0.003$, but $\sigma_{Pp}^{\rm SD}$ is essentially independent of $x_P$ for $0.001 \lapproxeq x_P \lapproxeq 0.01$, see Fig.~\ref{fig:A4}. $\sigma_{pp}^{\rm ND}$ is compared with a completely independent prediction (upper curve) obtained from $pp$ elastic (and diffractive) data \cite{LKMR}.}
\label{fig:A2}
\end{center}
\end{figure}

The resulting values of the cross section, $\sigma_{pp}^{\rm ND}$, are shown in Fig.~\ref{fig:A2}. Note that the true non-diffractive cross section, $\sigma_{pp}^{\rm ND}$, is much smaller once account is taken of the absorptive effects described by the multi-Pomeron diagrams. We see from Fig.~\ref{fig:A2} that the values of the one-Pomeron cross section are in broad agreement with the values obtained in Ref.~\cite{LKMR} based on the analysis of quite a different set of data, that is {\it elastic} $pp$ (and $\bar{p}p$) scattering data. Note that the inelastic cross section obtained from this elastic $pp$ analysis \cite{LKMR} also included the contributions from single- and double-diffractive dissociation of the protons. These diffractive processes are absent in the non-diffractive events generated by PYTHIA. Therefore it is not surprisingly that the LKMR \cite{LKMR} cross section is a bit larger than that obtained using (\ref{eq:sig}).

\begin{figure} 
\begin{center}
\includegraphics[height=6cm]{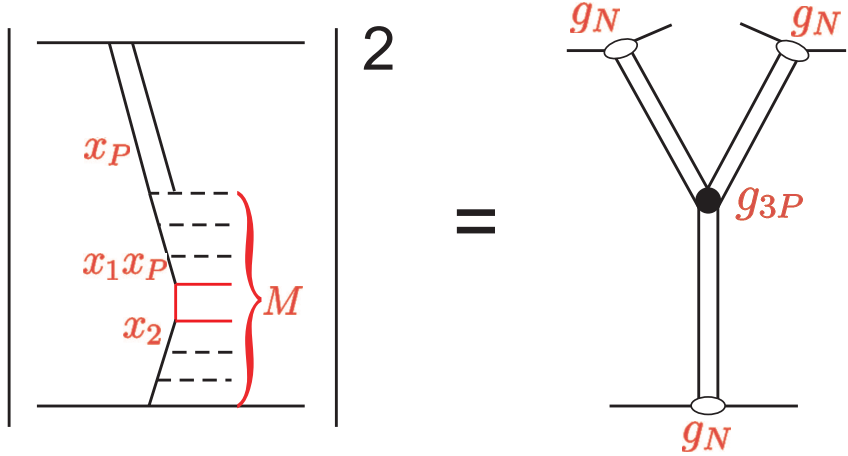}
\caption{\sf The triple-Pomeron diagram describing the  single-diffractive process in which the lower proton dissociates into a system of large mass $M$. Now $x_1 x_P ~(x_1)$ is the fraction of the upper proton's (Pomeron's) momentum carried by the active parton.}
\label{fig:A3}
\end{center}
\end{figure}

\begin{figure} 
\begin{center}
\includegraphics[height=8cm]{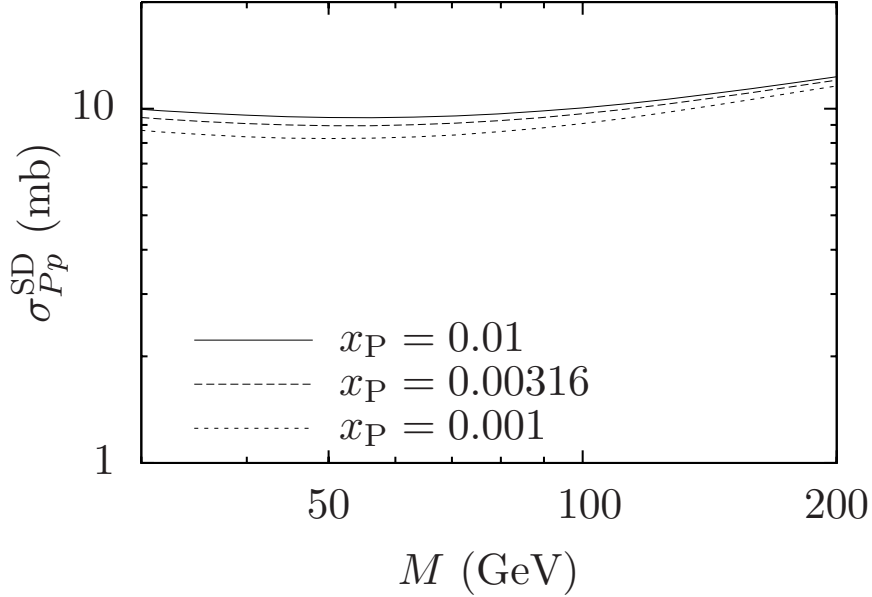}
\caption{\sf The dependence of $\sigma^{\rm SD}_{Pp}$ on the value of $x_P$. The three values chosen for $x_P$ span the region of the HERA diffractive data.}
\label{fig:A4}
\end{center}
\end{figure}

In Regge theory, the next step is to account for Pomeron-Pomeron interactions. The simplest example is the so-called triple-Pomeron diagram. It describes the process where the system of outgoing hadrons is separated from a beam proton by a large rapidity gap. The diagram is sketched in Fig.~\ref{fig:A3}. The size of the corresponding cross section (which we denote $\sigma_{Pp}^{\rm SD}$) is driven by the value of the triple-Pomeron vertex $g_{3P}$. Usually the value of $g_{3P}$ is determined by a global analysis of the data in the triple-Regge region, where the outgoing beam proton carries away a large fraction of its initial momentum, that is $x_L=1-x_P$ is close to 1.
It is possible, and desirable, to evaluate the important triple-Pomeron coupling, $g_{3P}$, by a qualitatively different procedure based on the known diffractive parton distribution functions (DPDFs).  The DPDFs are determined by analyses \cite{DPDF,MRW,H1} of deep inelastic events with large rapidity gaps that were observed at HERA. Essentially this alternative procedure is to replace one of the inclusive PDFs in (\ref{eq:sig}), say $g(x_1)$, by the known DPDF, $g^D(x_1)$, or, to be precise $g^D(x_1)/f_P(x_P)$, see below. In this way we can determine the value of the inelastic Pomeron-proton cross section, which can be written in the form\footnote{Only the Pomeron contribution to the DPDFs is used in the present calculations. The secondary Reggeon component corresponds to the $RRP$ vertex.}
\be
\sigma_{Pp}^{\rm SD}~=~g_{3P}g_N~\left(\frac{M^2}{s_0}\right)^{\alpha(0)},
\label{eq:s1}
\ee
which can then be compared with the $pp$ inelastic cross section
\be
\sigma_{pp}^{\rm ND}~=~g_N^2~\left(\frac{M^2}{s_0}\right)^{\alpha(0)}.
\label{eq:s2}
\ee
It is implicit that (\ref{eq:s1}) and (\ref{eq:s2}) refer to the one-Pomeron exchange cross sections. The mass of the final hadron state is $M=\sqrt{x_P s}$ in the $Pp$ case and $M=\sqrt{s}$ for the $pp$ case. However, the DPDFs are not presented in terms of the parton distributions inside the Pomeron, but already include the Pomeron flux factor. That is, they already include the vertices and upper Pomeron propagators in Fig.~\ref{fig:A3}. Therefore to obtain $\sigma_{Pp}^{\rm SD}$  we have had to divide the result obtained from (\ref{eq:sig}) (with $g(x_1)$ replaced by $g^D(x_1)$) by the Pomeron flux factor
\be
f_P(x_P)~=~\frac{g^2_N}{16\pi^2 B}~(x_P)^{2(1-\alpha_P(0))}.
\ee
The values of the $t$-slope of the Pomeron-proton interaction, $B$, and the Pomeron trajectory, $\alpha_P$, were taken to be those of the DPDF analysis of Ref.~\cite{MRW}, while the Pomeron-nucleon coupling, $g_N$, is taken from Ref.~\cite{LKMR}. Strictly speaking the DPDFs depend on the value of $x_P$. However it is seen from Fig.~\ref{fig:A4} that the value of $\sigma_{Pp}^{\rm SD}$ is not strongly dependent on the value of $x_P$ in the kinematic range where diffractive deep inelastic data were measured at HERA, that is, $0.001 \lapproxeq x_P \lapproxeq 0.01$. This fact confirms the Regge factorization approach which is used in the analysis of diffractive deep inelastic data.

We note that the tuning of the PYTHIA Monte Carlo, with the parameters given in (\ref{eq:para}), was obtained using a leading order (LO) framework, while the DPDFs were obtained using a NLO analysis. There is a large difference between the LO and NLO inclusive PDFs for the proton, which arises mainly from the absence of the $1/z$ singularity in the LO $qq$ splitting function. In general, this difference is compensated by NLO corrections to the matrix elements, but since the only tuning of $p_{\rm min}$ of (\ref{eq:para}) was done within the LO framework we cannot use NLO proton distributions. 

\begin{figure} 
\begin{center}
\includegraphics[height=8cm]{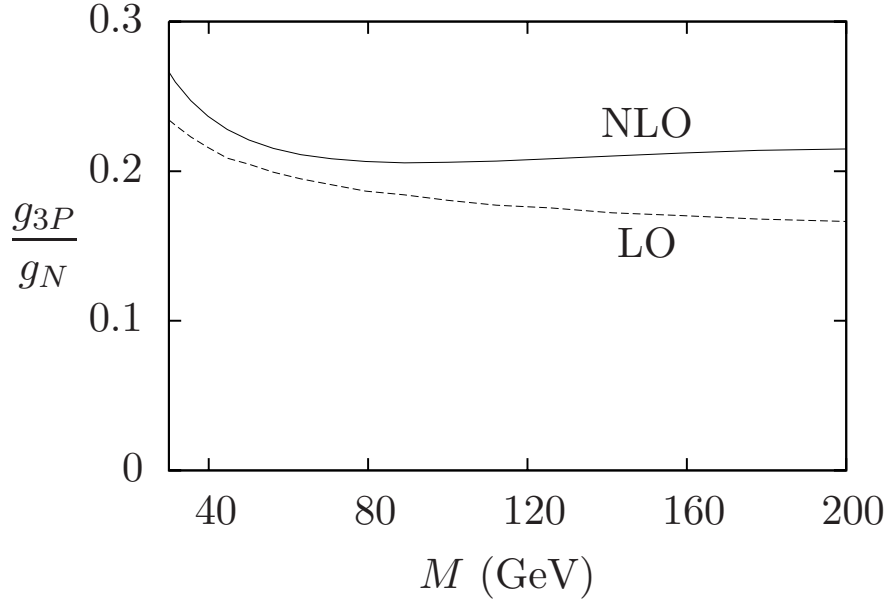}
\caption{\sf The value of the ratio $g_{3P}/g_N$ determined using DPDFs measured at HERA as a function of the mass $M$ of the system resulting from proton dissociation. The NLO curves were obtained using the DPDFs of \cite{MRW}, whereas the LO curves result from the DPDFs of the H1 analysis \cite{H1}.}
\label{fig:A5}
\end{center}
\end{figure}

In the diffractive case (the DPDFs of the Pomeron) the gluons dominate the input distributions; thus LO and NLO are reasonably close to each other. Therefore the calculation of $\sigma_{Pp}^{\rm SD}$ using the same LO CTEQ5L partons\footnote{Recall that these partons were used to tune the PYTHIA Monte Carlo.} as in th $pp$ case, and the NLO partons \cite{MRW} from the Pomeron side, should not cause a problem. Indeed, using DPDFs obtained at LO\footnote{These LO DPDFs based on H1 data \cite{H1} were extracted from the PYTHIA-8.1 code \cite{Pythia}.} in the H1 B fit of Ref.~\cite{H1}, we obtain a rather similar result for $\sigma_{Pp}^{\rm SD}$, as indicated by the dashed line in Fig.~\ref{fig:A5}. The difference between the two results can be considered as a measure of the uncertainty in the determination of the ratio $g_{3P}/g_N$. 

The ratio of the $Pp$ to the $pp$ cross sections in the relevant mass range, $30<M<200$ GeV are shown in Fig.~\ref{fig:A5}. The result is
\be
\frac{\sigma_{Pp}}{\sigma_{pp}}~=~\frac{g_{3P}}{g_N}~\simeq~0.2.
\ee
This estimate is in remarkable agreement with the value 0.2 obtained from the analysis \cite{LKMR} of pure soft {\it triple-Regge data}, that is from a completely different set of data to the {\it diffractive deep inelastic data} used for the present estimate. The growth of the `diffractive' cross section, $\sigma^{\rm SD}_{Pp}$, and of the ratio $g_{3P}/g_N$ as $M$ decreases below 50 GeV, is explained in Regge theory by the admixture of secondary Reggeons in the $Pp$-amplitude (namely the $PPR$ contribution), which is hidden in the `large' $x_1$ behaviour of the diffractive PDFs.

\section*{Acknowledgements}
We thank Valery Khoze for useful discussions. MGR would like to thank the IPPP at the University of Durham for hospitality. EGdeO is supported by CNPq (Brazil) under contract 201854/2009-0, and MGR is supported by the grant RFBR
11-02-00120-a, and by the Federal Program of the Russian State RSGSS-3628.2008.2.

\end{document}